\documentclass[aps,showpacs,floatfix,twocolumn,superscriptaddress]{revtex4}
\usepackage{amsmath}
\usepackage{graphicx}
\usepackage{epsfig}
\usepackage{lscape}
\usepackage{ulem}
\def\beq{\begin{equation}}
\def\eeq{\end{equation}}
\def\beqa{\begin{eqnarray}}
\def\eeqa{\end{eqnarray}}
\def\ban{\begin{eqnarray*}}
\def\ean{\end{eqnarray*}}
\def\bi{\begin{itemize}}
\def\ei{\end{itemize}}

\begin{document}
\title{Effects of the symmetry energy on the kaon condensates in the QMC Model}
\author{Prafulla K. Panda}
\affiliation{Department of Physics, C.V. Raman College of Engneering,
Vidya Nagar, Bhubaneswar-752054, India}
\affiliation{Centro de F\'{\i}sica Computacional - Departamento de F\'{\i}sica,
Universidade de Coimbra - P-3004 - 516 - Coimbra - Portugal}
\affiliation{Departamento de F\'isica, CFM, Universidade Federal de Santa 
Catarina, CP 476, CEP 88.040-900 Florian\'opolis - SC - Brazil}
\author{D\'ebora P. Menezes}
\affiliation{Departamento de F\'isica, CFM, Universidade Federal de Santa 
Catarina, CP 476, CEP 88.040-900 Florian\'opolis - SC - Brazil}
\affiliation{Departamento de F\'isica Aplicada, Universidad de Alicante, 
Ap. Correus 99, E-03080, Alicante, Spain}
\author{Constan\c ca Provid\^encia}
\affiliation{Centro de F\'{\i}sica Computacional - Departamento de F\'{\i}sica,
Universidade de Coimbra - P-3004 - 516 - Coimbra - Portugal}
\begin{abstract}
In this work we investigate protoneutron star properties within a
modified version of the quark coupling model (QMC) that incorporates a
$\omega-\rho$ interaction plus kaon condensed matter at finite temperature. Fixed entropy and
trapped neutrinos are taken into account. Our results are compared with the ones
obtained with the GM1 parametrization of the non-linear Walecka model
for similar values of the symmetry energy slope.  
Contrary to GM1, within the QMC the formation of low mass black-holes
during cooling are not probable. It is shown that the evolution of the
protoneutron star may include the melting of the kaon condensate
driven by the neutrino diffusion, followed by the formation of a
second condensate after cooling.  The signature of this complex
proccess could be a neutrino signal followed by a gamma ray burst.
We have seen that both models can, in general, describe very massive stars. 

\end{abstract}
\maketitle
\section{Introduction}

In the last years, all sorts of phenomenological equations of state (EoS), 
relativistic and
non-relativistic ones, have been used to describe (proto)neutron star matter.
These EoS are parameter dependent and 
are adjusted so as to reproduce nuclear matter bulk properties, as the
binding energy at the correct saturation density and incompressibility
as well as ground state properties of some nuclei and their 
collective responses \cite{fitting,nl3}. Attempts to constrain the EoS
have been made and they were based either on finite nuclei
experimental results, as for instance, the isoscalar monopole and the isovector
dipole giant resonances \cite{jorge2002} and neutron skin thickness
\cite{peles} or on astrophysical observations \cite{cottam,sanwal}.
Until 2010, when a star with a mass of almost 2 $M_\odot$ was
confirmed \cite{demorest}, most EoS  were expected to produce maximum stellar 
masses just larger than 1.44 $M_\odot$ and radii of the order of 10 to 12 km. 
Some of them, as the NL3 parametrization \cite{nl3} of the non-linear Walecka
model (NLWM) were even discarded to be considered too hard and to provide
too large solar masses. Recently
a second very massive neutron star was detected \cite{antoniadis} and many 
parametrizations and models were revisited and readjusted to account for 
the new observations. Also, many other constraints based on the above
mentioned nuclear properties and also on the symmetry energy, its
slope, skewness, dipole polarizabilities,
heavy-ion collision flows, isobaric analog states, etc, have been proposed
\cite{lattimer, outros}. 

As far as relativistic models are concerned, the $\omega-\rho$
interaction \cite{antigos,IUFSU} can be 
adjusted to reproduce experimental values of 
the symmetry energy and its related slope, the latest being
strongly correlated with many nuclear \cite{review} and stellar
properties \cite{rafael}.

On the other hand, it is well known that the EoS and the 
internal constitution of the
neutron stars depend on the nature of the strong interaction. 
In a compact star, strangeness may occur in the form of 
baryons, such as $\Lambda$ and $\Sigma^-$
hyperons, as a Bose condensate i.e. $K^-$ meson condensate, or in the
form of strange quarks, in all cases influencing the star structure
and its macroscopic properties \cite{prak97,Glen00}.  Some years ago, it was
suggested that above some critical density, the
ground state of baryonic matter might contain a Bose-Einstein
condensate of negatively charged kaons \cite{kaplan} as far as a
pion condensate also exists. In \cite{brown}, another mechanism that
allowed kaon condensation without pion condensation was proposed for
the first time. There is a strong attraction
between $K^-$ mesons and baryons which increases with density and
lowers the energy of the zero momentum state. A condensate is formed
when this energy equals the kaon chemical potential $\mu$.
When the electron chemical potential equals the effective kaon mass,
the kaons are favored because they help in the conservation of charge neutrality
once they are bosons and can condense in the lowest energy state. For this
reason $K^-$ mesons that have the same electric charge as electrons, 
 are the type of kaons that normally appear in a condensed state 
 in stars.

It is well known that the onset of the kaon condensation is
model dependent and varies according to the strength of the kaon 
optical potential \cite{kaon,gupta2012}.
In the present work, we revisit the possibility that a hybrid compact
star can be constituted by hadrons and  kaon condensed matter at
higher densities \cite{kaon}  by using the quark meson coupling
(QMC) model at finite temperature \cite{qmc,panda2010} with the inclusion of
the $\omega-\rho$ interaction \cite{antigos,IUFSU}. The inclusion of 
this non-linear interaction was shown to soften the symmetry energy 
at high densities and to bring the QMC model properties closer to density 
dependent relativistic models \cite{panda2012}. As the inclusion of this term
allows us to tune the slope of the symmetry energy, shown to be strongly 
correlated with some star properties, it is an important ingredient in the
investigation that follows. In a previous work \cite{gupta2013} a  discussion on
the onset of kaons and antikaons controlled by stiff and soft
symmetry energy and EoS was performed at zero temperature  with four
kinds of models: a standard relativistic mean field one, a density
dependent model, a model with the $\omega-\rho$ interaction and a
model with higher order coupling constants. It was found that
although the last two models bear quite different symmetry energies,
they yield very similar star  masses and radii. 
In \cite{pons2001}, it was seen that the effects of kaon condensation
on metastable stars can be quite dramatic resulting in different
neutrino emission signals. Hence, for a better
understanding of the role played by kaons inside a star, we use
two different models, the QMC and the GM1 \cite{Glen00} parametrization of the
NLWM for three values of fixed entropies that correspond to different
snapshots of the star evolution and discuss the effect of the
symmetry energy. The properties of the stars are also studied.

In the QMC model, nucleons are described 
as a system of nonoverlapping MIT bags that interact through the scalar 
and vector mesons. The quark degrees of freedom are explicitly taken 
into account and the couplings are determined at the quark level. 
In the present work we also treat the kaons as MIT bags \cite{bag} and the 
couplings of the kaons with
nucleons are determined in a self-consistent way.

In section II, we discuss the formalism employed in finite
temperature calculations. In section III we present and discuss our
results and compare them with the GM1 parametrization of the
non-linear Walecka model \cite{Glen00}.  For this comparison, in view
of the fact that the symmetry energy slope regulates some physical
properties, it was chosen to be similar in both models.
In the final section, we draw our conclusions.

\section{Formalism}
In the QMC model, the nucleon in nuclear medium is
assumed to be a static spherical MIT bag in which quarks
interact with the scalar $(\sigma)$ and vector $(\omega, \rho)$ fields, 
and those are treated as classical fields in the mean-field approximation
(MFA). The quark ${\psi_q}(\vec{r},t)$ inside the bag 
satisfies the equation of motion:

\begin{equation}
[i\gamma_\mu \partial^\mu - (m^0_q - g^q_\sigma \sigma_0) - 
\gamma^0 (g^q_\omega \omega_0 + 
\frac{1}{2} g^q_\rho \tau_{3q} b_{03} )] {\psi_q}(\vec{r},t)  =  0,
\end{equation}
where ${\sigma}_0 ,{\omega}_0, b_{03} $ are the classical meson fields for 
$\sigma,~\omega,~\rho$ mesons, ${m^0}_q$ is the current 
quark mass and $ {\tau}_{3q}$ is the third component of the Pauli 
matrices. $ g^q_\sigma,~g^q_\omega,~g^q_\rho$ denote the
quark coupling constants with $\sigma,~\omega,~\rho$.    
At finite temperature, quarks inside the bag can be thermally excited 
to higher angular momentum states and also  quark – anti-quark pairs can 
be created. For simplicity, the bag is assumed to be spherical with 
radius $R$ which depends on the temperature. The single particle energies 
in units of $R^{-1}$ for the quarks and the anti-quarks are given as 
\[ \epsilon_q^{nk}= \Omega_q^{nk} + R_N(V_\omega \pm \frac{1}{2} V_\rho)\]
and
\begin{equation} 
\epsilon_{\bar q}^{nk}=\Omega_q^{nk} - R_N(V_\omega \pm \frac{1}{2} V_\rho)
\end{equation} 
where $V_\sigma=g_\sigma^q\sigma_0$, $V_\omega=g_\omega^q\omega_0$ and
$ V_\rho=g_\rho^q b_{03}$. The $+ve$ sign refers to the $u-$quarks and
the $-ve$ sign to the $d-$quarks.
The total energy from the quarks and anti-quarks at finite temperature is 
\begin{equation} 
E_{tot}=\sum_{q,n,k}{\frac{\Omega_q^{nk}}{R_N}(f^q_{nk} + f^{\bar q}_{nk})}
\end{equation} 
where 
\[f^q_{nk} =  \frac{1}{e^{[ \Omega_q^{nk}/R_N - \upsilon_q]/T} + 1 },\]
and
\begin{equation} 
f^{\bar q}_{nk}=\frac{1}{e^{[ \Omega_q^{nk}/R_N + \upsilon_q]/T} + 1 } 
\end{equation}
with $\Omega_q^{nk}  =  \sqrt{x_{nk}^2 + R_N^2 m_q^{*2}}$ and 
the eigenvalues $ x_{nk} $ for the state  characterized by $n$ and 
$k$ are determined by the boundary condition at the bag surface.
In the above, $ \upsilon_q=\mu_q - V_\omega -m_\tau^q V_\rho$ is the 
effective quark chemical potential and 
is related to the quark chemical potential, $\mu_q$.
The energy of a static baryon bag consisting of three ground state quarks 
can be expressed as 
\begin{equation}
E_N^{bag} =  E_{tot} - \frac{Z_N}{R_N} + \frac{4}{3} \pi {R_N}^3 B_N
\end{equation} 
where  $Z_N $ is the parameter that accounts for zero point motion and 
$B_N $ is the bag constant. The entropy of the bag is defined as,
\begin{eqnarray}
S^{bag} &=& -\sum_{q,n,k} [f^q_{nk} \ln{f^q_{nk}}+
(1-f^q_{nk})\ln{(1-f^q_{nk})}\nonumber\\
&+& {\bar f}^q_{nk}\ln{{\bar f}^q_{nk}}+(1-{\bar f}^q_{nk})
\ln{(1-{\bar f}^q}_{nk})].
\end{eqnarray} 
The free energy of the bag is given by
$F_N^{bag} = E_N^{bag} - T S^{bag}$ and
the effective mass of a nucleon bag at rest is taken to be 
$M_N^* = F_N^{bag}$.
The equilibrium condition for the bag is obtained by minimizing the 
effective mass ${M_N}^*$ with 
respect to the bag radius $R_N$   
\begin{equation}
\frac{\partial{M_N^*}}{\partial{R_N^*}} = 0.
\end{equation} 
Once the bag radius is obtained, the effective baryon mass is immediately 
determined. For a given temperature and scalar field $\sigma$, the 
effective quark chemical potentials $\nu_q$ are determined from the 
total number of quarks, isospin density and strangeness.

We next obtain the thermodynamic potential within the mean field 
approximation and perform a calculation similar to that carried out 
in \cite{kaon}.
We assume that the kaons are described by the static MIT bag 
in the same way as the nucleons.
Moreover $\sigma,~\omega$ and $ \rho$ mesons are only mediators of 
the $u$ and $d$ quarks inside the kaons. The effective 
Lagrangian density for the kaon sector is
\begin{equation}
\mathcal{L}_K = D_\mu^* K^* D^\mu K - M_K^* K^* K
\end{equation}  
where kaons are coupled to the meson fields via minimal coupling and 
the covariant derivative reads
\begin{equation}
D_\mu = \partial_\mu+i g_{\omega K}\omega_\mu+i \frac{1}{2}g_{\rho K} 
\vec{\tau} \cdot b_\mu ,
\end{equation}
with
\begin{equation}
X_\mu=g_{\omega K}\omega_\mu+g_{\rho K}\vec\tau\cdot\vec\rho_\mu.
\end{equation}
The energy of the static bag describing kaon K can be expressed as 
\begin{equation}
E_K^{bag} = \sum_{q,n,k}\frac{\Omega^{nk}_q}{R_K}
(f^q_{nk}+f^{\bar q}_{nk})-\frac{Z_K}{R_K}+\frac{4}{3}\pi R_K^3 B_K.
\end{equation} 
For our calculations, we have fixed the bag constant, $B_K$, to be the
same as for the nucleon and from the kaon mass and the stability condition
in the vacuum, we have obtained $Z_K=3.362$ and $R_K=0.457$ fm for $R_N=0.6$ fm.
The free energy of the kaon bag is
$F_K^{bag} = E_K^{bag} - T S^{bag}$
and the effective mass of a kaon bag at rest is taken to be 
$ M_K^* =  F_K^{bag}$. In analogy with the nucleonic sector, the equilibrium condition for the bag is
\begin{equation}
\frac{\partial{M_K^*}}{\partial{R_K^*}} = 0.
\end{equation} 
The Bose occupation probability for particles $(f_{B^+})$ and 
anti-particles $(f_{B^-})$ appears naturally in the equation of state and reads
\begin{eqnarray}
f_{B^\pm}&=&\frac{1}{(e^{\beta (\omega^\pm\mp\mu_K)}-1)}
\end{eqnarray}
with $\beta=1/T$,
$\epsilon_K^*=\sqrt{p^2+{M_K^*}^2}$ and $\omega^\pm=\epsilon_K^*\pm X_0$.
In the above we define the kaon effective chemical potential $\nu_K=\mu_K+X_0$
where $X_0=g_{\omega K}\omega_0+g_{\rho K}b_{03}$. 

In the mean field approximation, the kaon contribution to 
the thermodynamic potential is
\begin{eqnarray}
&&\frac{\Omega_K}{V}=\zeta^2 \Big[{M_K^*}^2-(\mu_K+X_0)^2\Big] \nonumber\\
&+&T\int_0^\infty\frac{d^3p}{(2\pi)^3}\Big\{\ln[1-e^{-\beta(\omega^-+\mu)}]
+\ln[1-e^{-\beta(\omega^+-\mu)}]\Big\},\nonumber\\
\end{eqnarray}
from which we get
\begin{eqnarray}
P_K&=&-\frac{\Omega_K}{V}=
 \zeta^2 \Big[(\mu_K+X_0)^2-{M_K^*}^2\Big] \\
&+&\frac{1}{3}\int_0^\infty\frac{d^3p}{(2\pi)^3}\frac{p^2}
{\epsilon_K^*}\Big[f_{B^+} +f_{B^-}\Big],
\end{eqnarray}
and the kaon contribution to the energy density reads
\begin{eqnarray}
\varepsilon_K &=& \zeta^2 \Big[{M_K^*}^2+(\mu_K^2+X_0^2)\Big] \\
&+ &\int_0^\infty\frac{d^3p}{(2\pi)^3}
\Big[\omega^+(p)f_{B^+}+\omega^-(p)f_{B^-}\Big]\nonumber\\
&\equiv& \zeta^2 \Big[{M_K^*}^2+(\mu_K^2+X_0^2)\Big] \nonumber \\
&+& \int_0^\infty\frac{d^3p}
{(2\pi)^3}\epsilon_K^*\Big[f_{B^+} +f_{B^-}\Big].
\end{eqnarray}
The kaon number density is
\begin{equation}
n_K=n_K^c+n_K^{th},
\end{equation}
where $n_K^c= 2\zeta^2(\mu_K+X_0)$  is the condensate contribution and 
$n_K^{th}$ is the thermal contribution for the number density given by
\begin{equation}
n_K^{th}=\int_0^\infty\frac{d^3p}{(2\pi)^3}\Big[f_{B^+} -f_{B^-}\Big].
\end{equation}
Similarly the scalar density for the kaons is given by
\begin{equation}
n_K^s=\int_0^\infty\frac{d^3p}{(2\pi)^3}\frac{M_K^*}{\epsilon_K^*}
\Big[f_{B^+} +f_{B^-}\Big].
\end{equation}
The kaon entropy density is given by
$S_K=\beta(\varepsilon_K+P_K-\mu_K n_K)$.

The equations of motion for the meson fields are given by \cite{pmp}
\begin{eqnarray}     
{m_\sigma}^2 \sigma &=& \sum_{i=p,n}- \frac{\partial M_N^*}{\partial \sigma}
\nonumber\\ 
&\times&\frac{1} {\pi^2} \int d{\mathbf k}\frac{M_N^*} 
{\left[k^2 + M_N^{*2}\right]^{1/2}} (f_i+\bar f_i), \nonumber \\
&+& g_{\sigma K}(n_K^c+n_K^s)\\
{m_\omega}^2 \omega_0 &=& \sum_{i=p,n} {g_{\omega} \rho_i}-g_{\omega K} n_K
-2\Lambda_vg_\omega^2 g_\rho^2 b_{03}^2\omega_0,\\
{m_\rho}^2b_{03} &=&\sum_{i=p,n}{g_{\rho} I_{3i}\rho_i}-
\frac{1}{2}g_{\rho K} n_K\nonumber\\
&-& 2\Lambda_vg_\omega ^2 g_\rho^2 b_{03}\omega_0^2,
\end{eqnarray}    
and
\begin{equation}     
\zeta\Big[\mu_k-\omega^+(0)\Big]\Big[\mu_k+\omega^-(0)\Big]=0,
\end{equation}    
where
$f_i$ and $ \bar{f_i}$ are the thermal distribution functions 
for the baryon and antibaryon:
\begin{equation}
f_i =\frac{1}{e^{(\epsilon^* - \upsilon)/T} + 1 }~~~~~~\mbox{and}~~~~~~
\bar f_i  =  \frac{1}{e^{(\epsilon^* + \upsilon)/T} + 1 }.
\end{equation}
$\epsilon^* = \sqrt{\vec{k}^2 + M_N^{*2}}$,
is the effective nucleon energy, and $\upsilon = \mu_N - g_\omega\omega
-I_{3i} g_\rho b_{03} $ is the effective nucleon chemical potential. 
The term $\Lambda_v ~g_\omega^2 ~g_\rho^2 ~b_{03}^2 ~\omega_0^2$
accounts for the $\omega-\rho$ interaction as proposed in
\cite{antigos,IUFSU} and already considered in \cite{panda2012}.

As for the parameters, we have used \cite{pmp} $g_\sigma^q=5.957$,
$g_{\sigma}=3g_\sigma^q S_N(0)=8.58$, 
$g_{\omega}=8.981$, $g_{\rho}=8.651$ with $g_{\omega} = 3g_\omega^q$ 
and $g_{\rho} =
g_\rho^q$. We have taken the standard values for the meson masses,
$m_\sigma=550$ MeV, $m_\omega=783$ MeV and $m_\rho=770$ MeV.
Note that the $s$-quark is unaffected by the $\sigma$,
$\omega$  and $\rho$ mesons i.e. $g_\sigma^s=g_\omega^s=g_\rho^s=0\ .$ The kaon
couplings are given by $g_{\omega K}=\frac{1}{3}g_{\omega}$, $g_{\rho
K}=g_{\rho}$ as in \cite{Glen00}.

After a self-consistent calculation, the kaon effective mass, $m^*_K$
can be parametrized as \cite{tsushima98}
\begin{equation}
m^*_K=m_K-g_{\sigma K}(\sigma) \sigma \simeq m_K-\frac{1}{3} g_{\sigma}
\left(1-\frac{a_K}{2}g_{\sigma} \sigma \right) \sigma,
\end{equation}
where $a_k=0.00045043\mbox{ MeV}^{-1}$ for $R_N=0.6$ fm. This determines
the $g_{\sigma K}$ which is a density dependent parameter.

Finally, the total energy density of the nuclear matter with kaons at finite 
temperature becomes
\begin{equation}
\varepsilon = \varepsilon_B +\varepsilon_K, 
\end{equation} 
where \cite{panda2012}
\begin{eqnarray}
\varepsilon_B &=& \frac{2}{{(2 \pi)}^3} \sum_{i=p,n}\int d^3 k 
\left[\epsilon^* (f_i + \bar{f_i})\right]\nonumber\\
&+&\frac{1}{2} {m_\sigma}^2 \sigma^2 - \frac{1}{2} {m_\omega}^2 \omega_0^2 
- \frac{1}{2} {m_\rho}^2 b_{03}^2\nonumber\\
&+&g_\omega \omega_0\rho_B +\frac{1}{2}g_\rho ~b_{03}~\rho_3
-\Lambda_v ~g_\omega^2 ~g_\rho^2 ~b_{03}^2 ~\omega_0^2.
\end{eqnarray}

For neutron stars, their particle composition is determined by the requirements
of charge neutrality and $\beta$-equilibrium conditions under the
weak processes $n\Rightarrow p+l+\bar \nu$ and $n+l\Rightarrow p+\nu$,
implying that
$$\mu_n =\mu_p + \mu_e$$
and
\begin{equation}
\rho_e+\rho_\mu+\rho_K=\rho_p.
\end{equation}
If neutrino trapping
is imposed to the system, the beta equilibrium condition is altered to
$$\mu_n =\mu_p + (\mu_e-\mu_{\nu_e}).$$

In this work, different snapshots of the star evolution are simulated
through different entropies per particle and trapped neutrinos.
At first, the star is relatively warm (represented by fixed entropy per particle) and has a large number of
trapped neutrinos (represented by fixed lepton fraction). As the trapped 
neutrinos diffuse out, they heat up the star \cite{prak97}. Finally, 
the star is considered cold (zero temperature) and deleptonized:
\begin{itemize}
\item $S/\rho_B=1$, $Y_l=0.4$,
\item $S/\rho_B=2$, $\mu_{\nu_l}=0$,
\item $S/\rho_B=0$, $\mu_{\nu_l}=0$.
\end{itemize}
The last scenario has already been considered in \cite{kaon}. We next
discuss the first two cases and also the possibility (only for
academic purposes) that the temperature is fixed through out the star.
The number of muons and muon neutrinos is negligble when
the  fraction of electrons and electron neutrinos is fixed at 0.4 and,
therefore, in  this case we do not include muons in the calculation.
 For the cases without neutrinos, muons are also considered.

\section{Results and Discussion}

As mentioned in the Introduction, we discuss next the effect of the slope of 
the symmetry energy on the kaon condensation. The symmetry energy is defined as:
\begin{equation}
 {\cal E}_{sym} = \frac{1}{2} \left [ \frac{\partial^2 ({\varepsilon_B}/\rho_B)}{\partial \alpha^2}  \right ]_{\alpha = 0}=
\frac{k_F^2}{6 E_F}+\frac{g_\rho^2}{4{m^*_\rho}^2}\rho_B
\; ,
\end{equation}
where ${\varepsilon_B}$ is the energy density, 
$\alpha$ is the asymmetry parameter $\alpha  = (\rho_n - \rho_p)/\rho_B$,
$\rho_B=\rho_n+\rho_p$, $E_F=\sqrt{k_F^2+{M^*_0}^2}$, where 
$M_0^*$ is the nucleon effective mass at saturation density and 
$k_F=(3\pi^2 \rho_B/2)^{1/3}$, ${m^*_\rho}^2=m_\rho^2 +2\Lambda_v g_\omega^2 g_\rho^2
\omega_0^2$, 
and the slope of the symmetry energy is:
\begin{equation}
 L = \left [ 3 \rho_B \frac{\partial {\cal E}_{sym}}{\partial \rho_B}   \right ]_{\rho_B = \rho_0} \; .
\end{equation}
We consider six different EOS, obtained from the QMC and the GM1 
parametrization of
the NLWM. Both models have a similar symmetry energy and corresponding
slope at the saturation denstiy.  Including the  $\omega\rho$ term we build
other EOS from these models  by changing the symmetry energy slope at 
saturation and keeping
fixed the symmetry energy as 21.74 MeV at $\rho=0.1$ fm$^{-3}$. 
For each model, QMC and GM1, we have fixed  three different values of
the slope $L$,  roughly  the same  for both models, namely $L\sim 59.5,\, 70.5, \, 93.5$ MeV.
With a similar behavior of the symmetry energy at saturation, 
we can discuss how the other properties of the EOS affect the
properties of neutron stars containing a kaon condensate.
In table \ref{table0} we display the $L$ values and the related
$\Lambda_v$ and $g_\rho$ for the models we investigate. 

\begin{table}[t]
\centering
\renewcommand{\arraystretch}{1.4}
\setlength\tabcolsep{3pt}
\begin{tabular}{cccccc}
\hline
model& $L$ (MeV) & $ {\cal E}_{sym}$ (MeV) & $\rho_0$ (fm$^{-3}$)
& $\Lambda_v$ & $g_\rho$ \\
\hline
QMC & 93.5 &33.70 &  0.15  & 0.0  & 8.8606 \\
QMC & 70.5 &31.88 &  0.15  & 0.03 & 9.2463\\
QMC & 59.3 &30.87 &  0.15  & 0.05 & 9.5335\\
\hline
GM1 & 93.8 &32.47 & 0.153 & 0.0  &  8.0104 \\
GM1 & 70.8 &29.57 & 0.153 & 0.03 & 8.0104\\
GM1 & 59.6 &27.80 & 0.153 & 0.037 & 8.0104\\
\hline
\end{tabular}
\caption{\label {table0} Symmetry energy, its slope and related model 
parameters.}
\end{table}

For  the kaon-$\omega$ coupling we consider $g_{N\omega}/3$. In GM1, the coupling to
the scalar $\sigma$-field is fixed to a  kaon optical potential in symmetric
nuclear matter at saturation,  $V_K = - 125$ MeV, a value suggested by chiral
models \cite{chiral}. For this value of the kaon optical potential we obtain a
second order phase transition from a pure hadronic phase to a  hadronic phase
with kaons. A more attractive potential lowers the kaon onset density
and the transition to the hadronic phase is a first order one \cite{Glen00}.
Within  the QMC model  this quantitity is an output, and  with the present
choice of parameters  $V_K=
- 123$ MeV \cite{kaon} at saturation,  very close to the above value taken for GM1.

\begin{figure}[ht]
\includegraphics[width=0.8\linewidth,angle=0]{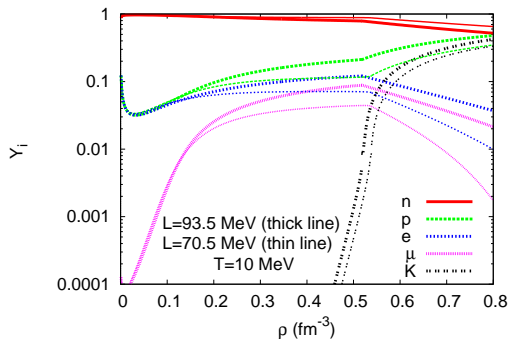}
\caption{Particle fractions obtained with the QMC model for T=10 MeV, L=93.5 MeV (thick line)
  and L=70.5 MeV (thin line).}
\label{fig1}
\end{figure}

In \cite{predeal2012} it was shown, for $T=0$ MeV, that: a) the larger $L$ the
larger the fraction of kaons at a given density; b) the onset of kaon
condensation occurs at  similar or slightly smaller densities for lower values of $L$.  It was also pointed out that the
EoS with the lower slope $L$ is softer, giving rise to stars with smaller
radii and with a larger total strangeness content because  of
 their  larger central densities. In the following we discuss the effect of
 the slope $L$ on the properties of warm stars with a kaon condensate described by
 two different models, QMC and GM1.

\begin{figure}[ht]
\includegraphics[width=0.8\linewidth,angle=0]{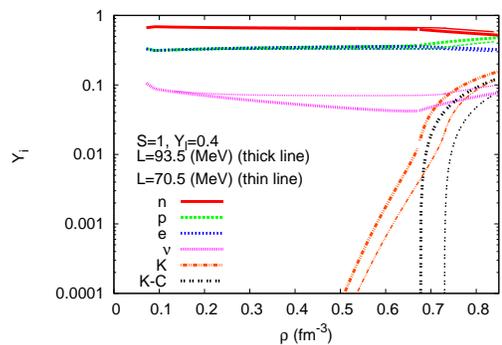}\\
\includegraphics[width=0.8\linewidth,angle=0]{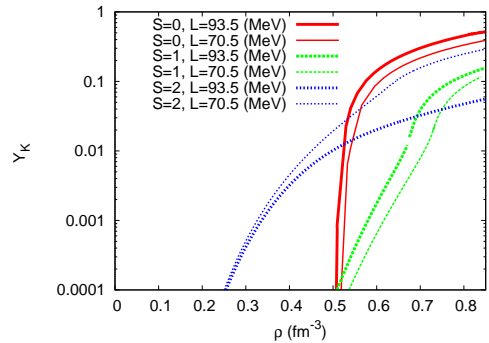}
\caption{Warm matter with trapped neutrino,  $S=1$ and $Y_l=0.4$,
a) particle fraction  versus the baryonic density within QMC  for  two different values of the slope $L$,; 
b) the kaon fraction
 for both QMC and GM1.}
\label{fig2}
\end{figure}

We first discuss the effect of the temperature on the kaon
  onset. The EOS is calculated at a fixed temperature, even though
 inside a compact object the temperature is not expected
to be constant.
In Fig. \ref{fig1} we plot the particle fraction for a fixed
temperature $T=10$ MeV and two different $L$ values within the QMC
model.  The temperature helps the
appearance of kaons (strangeness): they appear 
at a smaller density as compared will
the results presented in \cite{kaon} for a zero temperature
system. Similar results were obtained previously \cite{pons2001,banik}.
Moreover, it is clear that a larger slope enhances the
production of kaons. A larger $L$ favors larger proton and
  electron fractions, and, therefore we may also expect a larger kaon
  fraction, since kaons replace the  electrons.

\begin{figure}[ht]
\includegraphics[width=0.8\linewidth,angle=0]{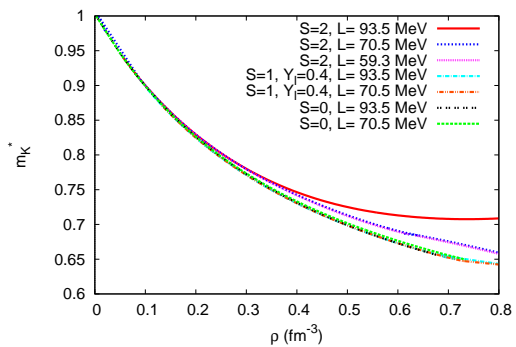}
\caption{Kaon effective mass  versus density in $\beta$-equilibrium
  matter  for different values of
  the entropy per baryon $S$  and the slope $L$. Matter for $S=1$
  contains trapped neutrions and all other curves were obtained for
  neutrino free matter.} 
\label{fig3}
\end{figure}

After a short initial time,  the entropy is pratically constant
inside the star, and therefore in the following we present results
obtained for a fixed entropy per baryon, for both matter with 
trapped neutrinos and $S=1$ and without neutrinos and $S=2$
\cite{prak97}.
In Fig. \ref{fig2}a), the results for kaon fractions obtained at a
fixed entropy per baryon $S=1$ and a lepton fraction of $Y_l= 0.4$ are
shown for the QMC and GM1 models and two different values of $L$.
The onset of a condensate of kaons occurs only at a density 
$\sim 0.15-0.2$ fm$^{-3}$ above the onset  of thermal kaons. 
Within GM1,  kaons
appear at lower densities and, for a given density, in larger amounts
than QMC, even when the same slope $L$ is chosen. Hence, stars descibed by 
the QMC model present a lower amount of strangeness.This behavior results from a
softer EOS at
densities above saturation within QMC. From Fig. \ref{fig2}b), it is  also seen that smaller values of
$L$ give rise to smaller amounts of strangeness, as already seen in 
 \cite{panda2012}, where hyperons (instead of kaons) were
 considered. The only exception is $S=2$: in this case the kaon
 condensation for $L=93$ MeV occurs at a density above the range of
 densities shown, and, therefore, the kaon fraction is always below the
 values obtained for $L=70$ MeV, which  predicts a kaon condensate
 above 0.6 fm$^{-3}$. The presence of neutrinos also disfavors the
 formation of kaons as seen if one compares the $S=1$ curves obtained with
 trapped neutrinos with the ones obtained with $S=0$ and $S=2$ neutrino free matter.
The total amount of kaons in the star is ultimately
 ditacted by the central density that is larger for the softer EOS. 
  
\begin{figure}[ht]
\includegraphics[width=0.8\linewidth,angle=0]{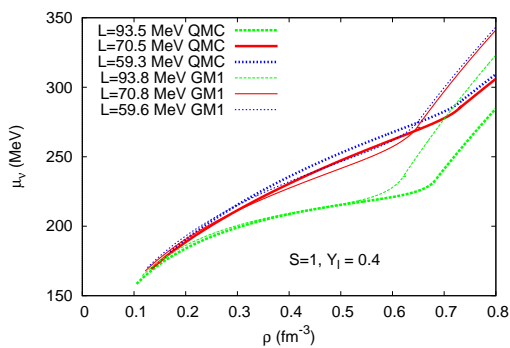}
\caption{Neutrino chemical potential versus baryon density for
  $\beta$-equilibrium matter at $S=1$ with the lepton fraction
  $Y_l=0.4$. Results for GM1 and QMC are shown for three different
  values of $L$.}
\label{fig5}
\end{figure}

 In Fig. \ref{fig3} the density dependence of the kaon effective mass is shown
for several values of the entropy per baryon and for different values of the
symmetry energy slope. Up to 2 times the saturation
density, the curves are practically identical, i.e., no dependence on the 
slope and temperature is noticed, but at high densities, the mass decreases 
faster for lower temperatures. In particular, for $S=2$ and $L=93$ MeV, the mass remains quite high and consequently, no
kaon condensate is formed, since a larger mass delays the onset of
kaons.

\begin{figure}[ht]
\includegraphics[width=0.8\linewidth,angle=0]{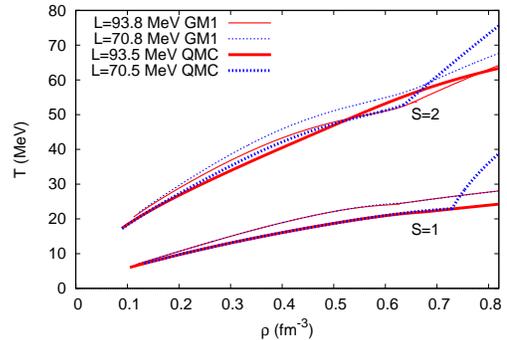}
\caption{Temperature versus density for neutrino free matter with
  $S=2$ and matter with neutrino trapped  matter with $S=1$.}
\label{fig6}
\end{figure}

\begin{table*}[t]
\centering
\renewcommand{\arraystretch}{1.4}
\setlength\tabcolsep{3pt}
\begin{tabular}{cccccccccc}
\hline
model& type&L (MeV) & $M_{max}(M_\odot)$& $M_{bmax}(M_\odot)$& R (km) &
$\varepsilon$ (fm$^4$)&$\varepsilon^K$ (fm$^4$)&$M^K_{max}(M_\odot)$
& $R(1.4 M_\odot)(Km)$\\
\hline
QMC& T=10 MeV&93.5&2.03 &2.37&12.8&5.66&2.80&1.86 & \\
QMC& S=1, $Y_l=0.4$ &93.5&2.03&2.25&11.7&5.26&4.09&1.99 & 13.12\\
QMC& S=1, $Y_l=0.4$ &70.5&2.05&2.19&11.1&5.97&4.41&1.98 & 12.56\\
QMC& S=1, $Y_l=0.4$ &59.03&2.04&2.25&10.9&5.97&4.30&1.97& 11.91\\
QMC& S=2 &93.5&2.51&3.11&12.2&5.35&3.03&2.30 & 14.15\\
QMC& S=2 &70.5&2.15&2.51&11.8&5.86&3.44&2.04 & 13.69\\
QMC& S=2 &59.03&2.13&2.48&11.7&5.99&3.57&2.03 & 13.54\\
QMC& S=0 &93.5&1.98&2.25&12.08&5.41&2.93&1.86 & 13.58\\
QMC& S=0 &70.5&1.94&2.12&11.85&5.61&2.96&1.81 & 13.19\\
QMC& S=0 &59.03&1.95&2.18&11.7&5.75&3.07&1.82 & 13.06\\
\\
GM1& S=1, $Y_l=0.4$ &93.8&2.24&2.52&11.9&5.52&3.70&2.17 &13.13\\
GM1& S=1, $Y_l=0.4$ &70.8&2.23&2.61&11.7&5.64&3.91&2.18 &12.96\\
GM1& S=1, $Y_l=0.4$ &59.6&2.24&2.53&11.7&5.64&3.92&2.18 &12.88\\
GM1& S=2 &93.8&2.31&2.65&12.7&4.88&3.97&2.30 &13.98\\
GM1& S=2 &70.8&2.26&2.60&12.3&5.11&4.09&2.25 &13.51\\
GM1& S=2 &59.6&2.27&2.60&12.2&5.13&4.41&2.25 &13.27\\
GM1& S=0 &93.8&2.14&2.46&12.8&4.63&2.70&2.03 & 13.82\\
GM1& S=0 &70.8&2.06&2.39&12.4&4.86&2.94&2.02 &13.23\\
GM1& S=0 &59.6&2.06&2.39&12.3&4.96&2.96&1.95 &13.06\\
\hline
\end{tabular}
\caption{\label {table1} Star properties for the EOSs described in the
  text for QMC and GM1 models and the three $L$ vaues:
the maximum gravitational and baryonic masses, corresponding
radius, the central energy density, onset energy density of the kaon
condensate, the mass of a star with this threshold mass 
and radii of $1.4M_\odot$ star.}
\end{table*}

In Fig. \ref{fig5} we plot the neutrino chemical potential as a
function of the baryonic chemical potential for $S=1$ and $Y_l=0.4$ 
for different values of the symmetry energy slope with QMC and GM1
models. Lower values of the slope correspond to larger amounts 
of neutrinos in matter with  a fixed fraction of leptons, because a smaller $L$ favors smaller amounts of protons and
  electrons at large densities. The kink at a chemical potential above
1300 MeV occurs at the onset of the kaon condensate.  After its onset,
the number of electrons decreases rapidly due to the charge
neutrality condition.
 For a fixed lepton the neutrino abundance increases to
compensate the decrease of the electrons.  Comparing the models, it is
clear that within GM1 the amount of neutrinos is larger, and therefore
a larger neutrino chemical potential is obtained.  We point out that
the kink in the chemical potential due to the onset of the condensate
is more pronounced for GM1 and occurs at lower densities. This is
due to the onset of a kaon condensation at lower densities and larger
amounts of condensed kaons at a given density. Therefore, we may
expect that during cooling a smaller amount of neutrinos is emmitted
within QMC and for smaller values of $L$, and, as a consequence we may
expect that the probability of the occurence of a black-hole will be
smaller within QMC. We come back to this point later.

\begin{figure}[ht]
\includegraphics[width=0.8\linewidth,angle=0]{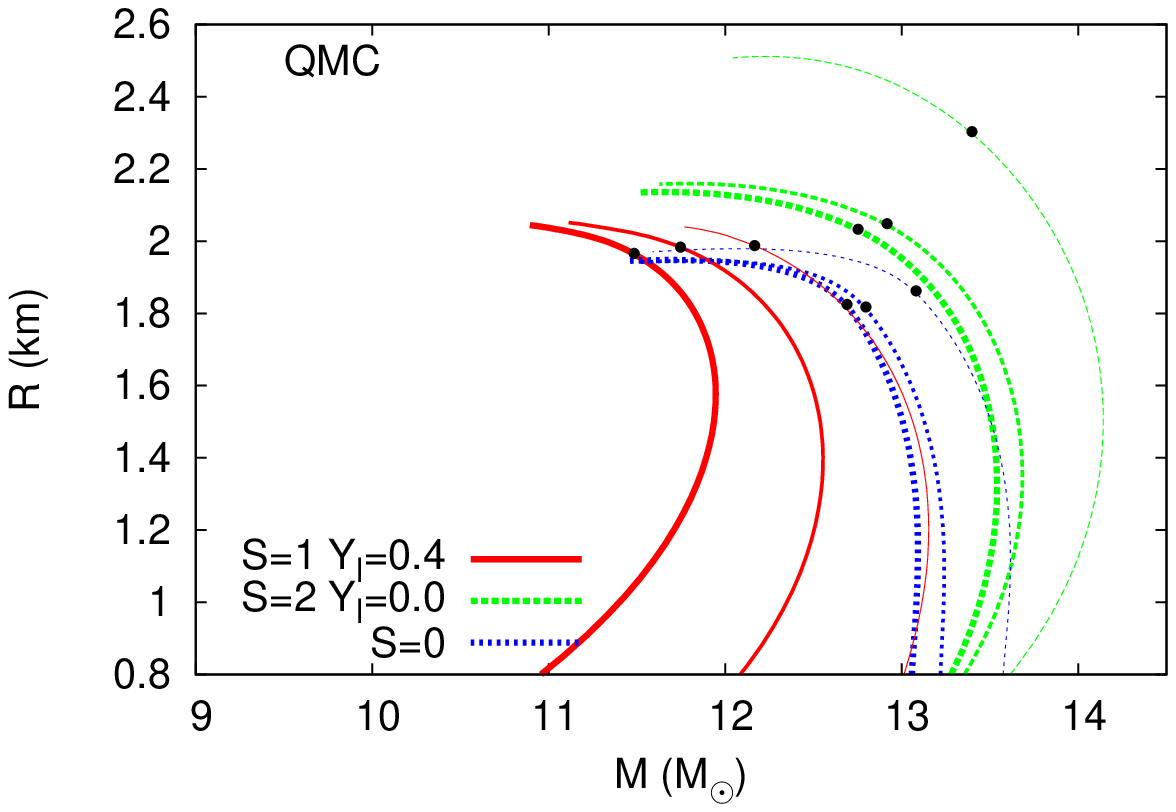}\\
\includegraphics[width=0.8\linewidth,angle=0]{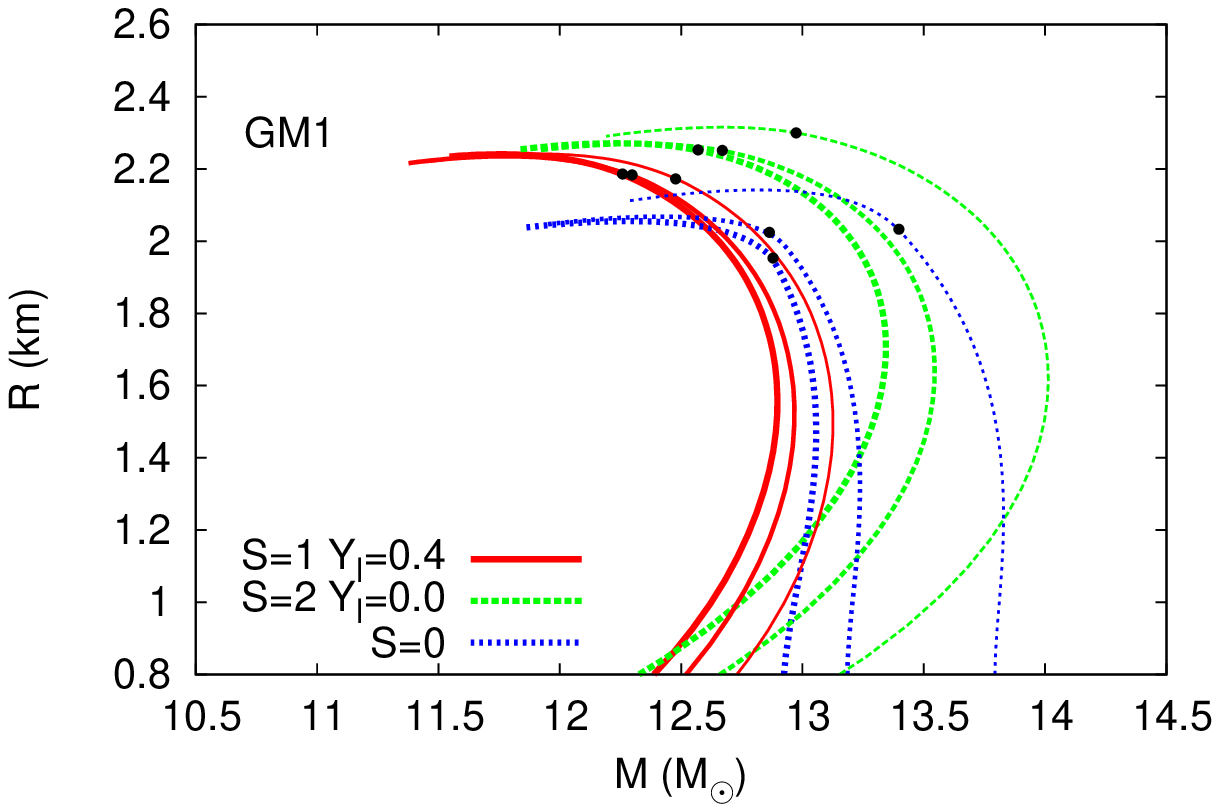}
\caption{Mass-radius profiles for stars with $S=1$ and fixed lepton
fraction $Y_l=0.4$ (red full lines), $S=2$ and neutrino free matter (green
dashed lines) and cold matter (blue dotted lines), for different 
values of the symmetry energy slope:
  $L=59$ (thick), 70 (medium) and  93 (thin) MeV, with  QMC
  (top panel) and GM1 (bottom panel). The black dots indicate the onset 
of kaon condensate}
\label{fig7}
\end{figure}

In Fig. \ref{fig6} the temperature of the system is depicted as a
function of the baryonic density for $S=2$ neutrino free
$\beta$-equilibrium matter  and $S=1$ $\beta$-equilibrium matter
with trapped neutrinos with both models under
investigation. As compared with the results shown in \cite{panda2010},
the system with kaons can reach temperatures inside
of the star that are much higher than the ones attained when hyperons
are included in its core, when the highest
temperature is 35 MeV. 
It is the kaon condensation the reason for
this behavior. It is seen that for $S=2$, $L=93$ MeV, the temperature
does not increase so much because the kaons do not condensate in the
range of densities shown.

\begin{figure*}[ht]
\begin{tabular}{cc}
\includegraphics[width=0.4\linewidth,angle=0]{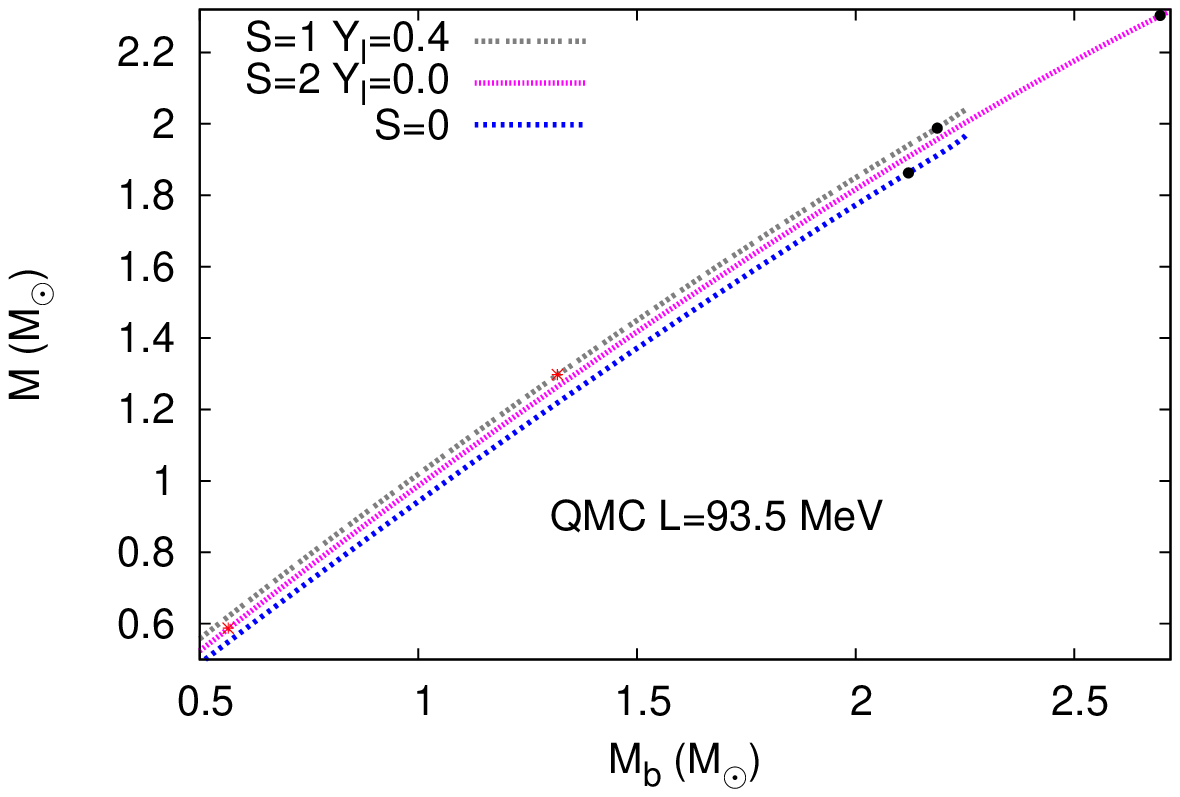}&
\includegraphics[width=0.4\linewidth,angle=0]{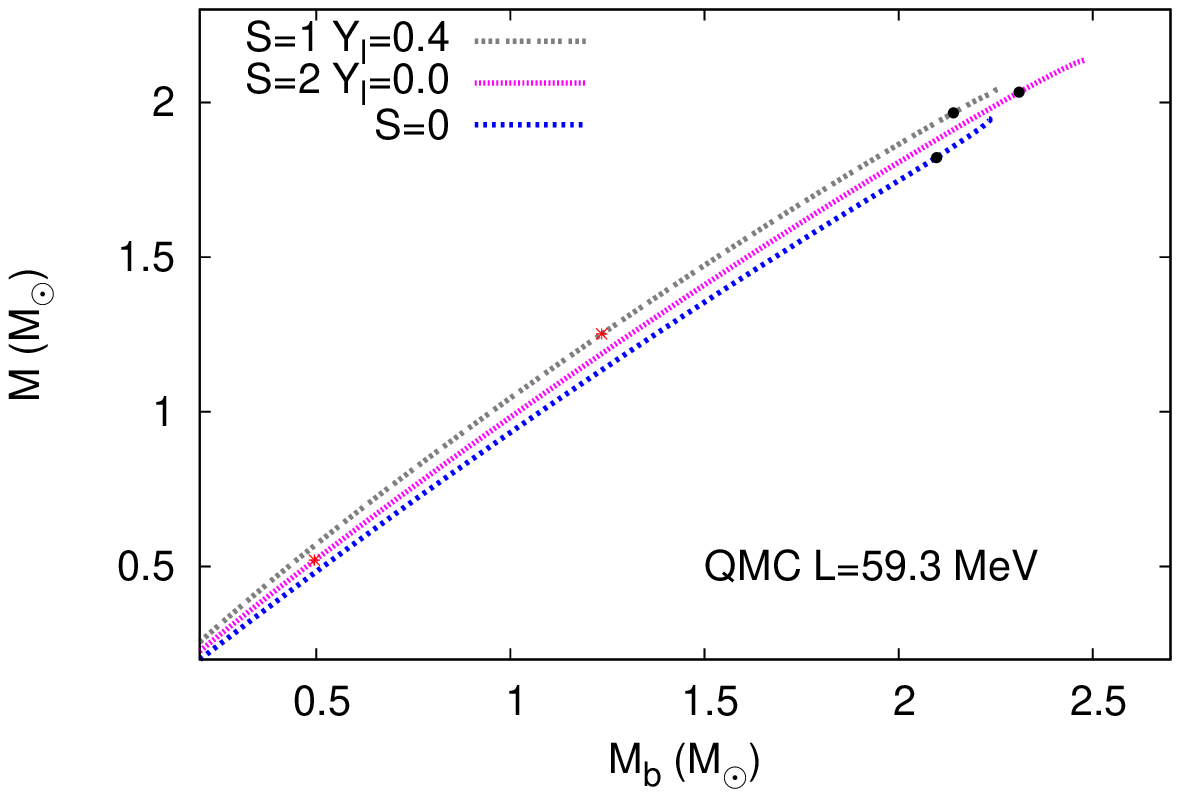}\\
\includegraphics[width=0.4\linewidth,angle=0]{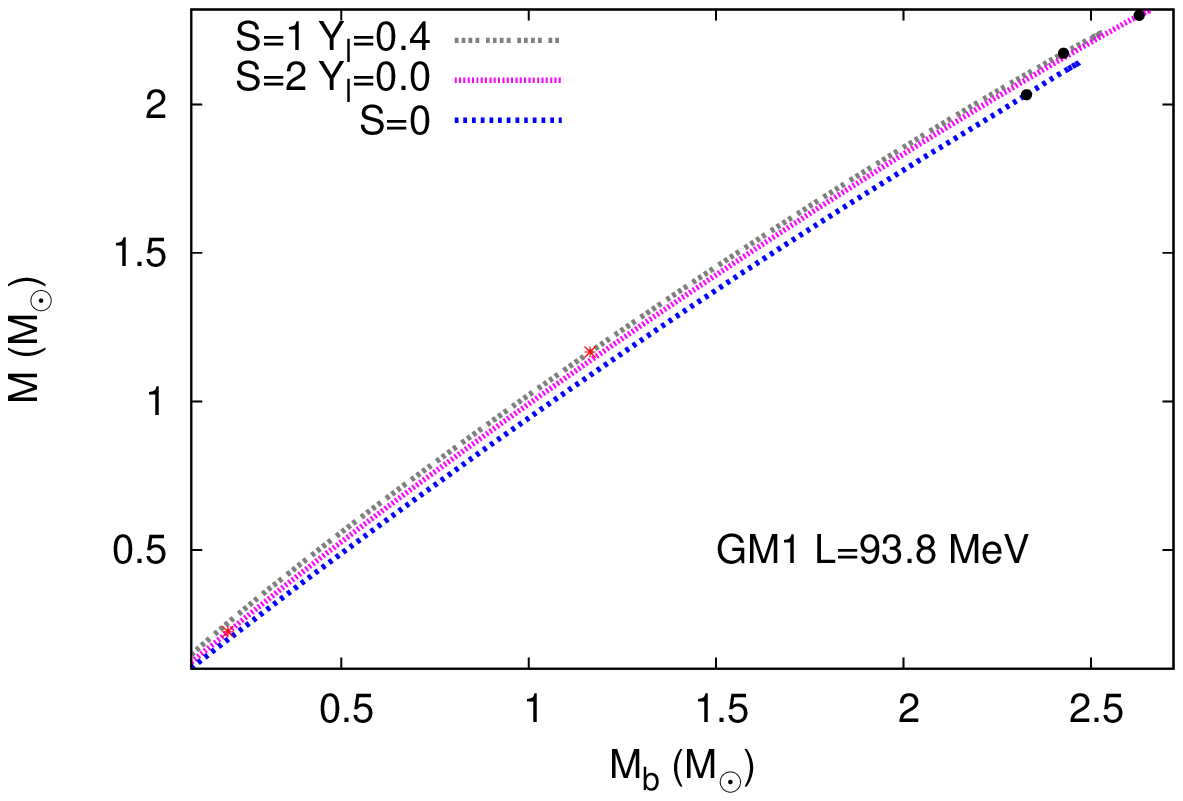}&
\includegraphics[width=0.4\linewidth,angle=0]{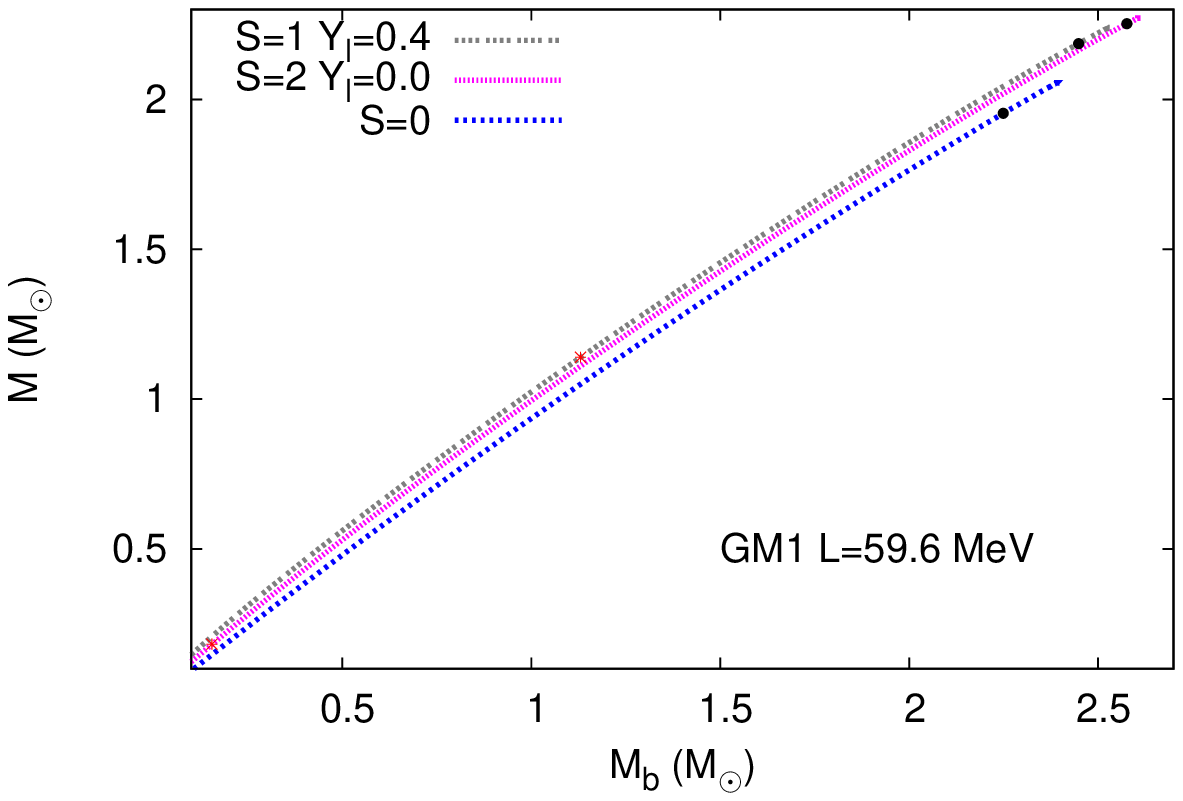}\\
\end{tabular}
\caption{Baryonic versus gravitacional masses for stars 
  with $S=1$ and fixed lepton
  fraction $Y_l=0.4$ (top curve), $S=2$ and neutrino free matter (middle
  curve) and cold matter (bottom curve) ,  for  $L=93.5$ MeV (left panels) and
  $L=59.3$ MeV (right panels). The results were obtained with QMC (top panels)
and GM1 (bottom panels). The black full dots indicate the minimal mass
configuration which contains a kaon condensate in the core, and the
red asterisks the minimal configuration with a fraction  $10^{-8}$ or
larger of thermal kaons.}
\label{fig8}\end{figure*}

We also see, from  Fig. \ref{fig6} that the 
temperature values are somewhat similar, but still slightly larger for
the smaller $L$ values if $S=2$. In matter with larger values of $L$ the number
of neutrons and protons is closer, which correspons to a smaller
temperature if we fix the entropy. No $L$  differences are seen in matter
with $S=1$, because in this case the entropy has an important contribution
from  neutrinos due to their almost zero mass.

Once the EoS are determined, we use them as input in the TOV equations
to obtain the stellar macroscopic properties, which are shown in
Table \ref{table1}  and Figs. \ref{fig7} and \ref{fig8}.
The result for $T=10$ MeV is added
only for completeness. The maximum masses for the cases with fixed
entropy do not show a clear behaviour with the slope, but for a stable
zero temperature system, they tend to decrease with the decrease of the slope
for both models. If the slope $L$ is small the trend may invert due to two
competing effects: a) a smaller slope $L$ means a softer EOS and, therefore,
smaller maximum masses; b) however the onset of kaons for a softer EOS occurs
at larger densities, as seen from table \ref{table1} from the central energy
densities of the threshold stars for the kaon onset. As a result stars
obtained from an EOS with larger $L$ contain larger amounts of kaons, which
soften the EOS.    Therefore, decreasing $L$ reduces the maximum star mass
until a critical value of $L$ when the onset of kaons is shifted to much
larger densities so that the overall softening due to the kaon fraction
becomes negligeable. $L=59$ MeV 
is one of these critical values of the
slope. This effect is present in both GM1 and QMC and for cold and warm
stars. A comment should also be done concerning the differences of maximum
masses with $L$. Within the QMC the maximum mass diference is quite small, in
fact not larger than 
$\sim 0.04$ M$_\odot$ except for the $S=2$, $L=93$ MeV case, when 
the much larger mass
is due to the non existing kaon condensation in the star, since only thermal
kaons are present. Within GM1 the differences are larger, but the main 
considerations above remain valid. The difference in this case is the fact that 
the nucleonic GM1 EOS is harder than QMC at intermediate densities.

The star radii, on the other hand,  decrease with
the decrease of the slope, see Fig. \ref{fig7}, as already seen in 
\cite{antigos,rafael} for
different parametrizations of the NLWM (excluding IU-FSU).
In these figures the black dots indicate the onset of thermal  kaons, which
occurs for quite massive stars.
Therefore, we may state that the radius difference between the families of
stars is mainly due to the different
behavior of the symmetry energy with density and not to the presence of the
kaons. According to \cite{Hebeler}, the radii of canonical $1.4M_\odot$ 
neutron stars should lie within the range 9.7-13.9 $Km$. In table \ref{table1}
we show our results for these canonical stars and see that, except for the
$S=2, L=93.5/93.8$ case, a star which is not stable, our results fall inside the
expected range. On the other hand, 
other two different analyses of five quiescent low-mass X-ray binaries in 
globular clusters were performed to establish possible neutron star radii
ranges. In the first analysis \cite{guillot}, all neutron stars 
were assumed to have the same radii in the range 
$R=9.1^{+1,3}_{-1.5}$. The second calculation \cite{Lattimer2013}, based on a 
Bayesian analysis, estimates that neutron stars radii should lie
in between 10.9 and 12.7 Km. It is important to have in mind that measurement 
and assessment of neutron star radii still remain to be better understood, but
if the above mentioned constraints are to be validated, our results would 
only be in accordance with the second analysis.
 
The kaon onset energy density increases with decreasing $L$ as alreay
referred, see $\epsilon^K$ in table \ref{table1}. However, since the existence
of kaons softens the EOS, the kaon threshold star mass is generally 
smaller for the smaller $L$. 
Notice that both models, even with the inclusion of kaons and the
$\omega-\rho$ interaction, known to
soften the EoS, may account for the description of very massive stars.

In Fig. \ref{fig8} we display, both for the QMC and GM1 models and two values 
of the symmetry energy slope, the baryonic versus
gravitacional masses for stars at different snapshots of their 
lives \cite{prakash01}: a)
immediatly after the core bounce with trapped neutrinos $S=1$ and fixed lepton
  fraction  $Y_l=0.4$; b) after neutrino diffusion $Y_\nu=0$  and core heating 
due to
  deleptonization. The maximum entropy per baryon  $S=2$ is attained at 
  $t\sim 15$ s; c) after core cooling with $S=0$ and   $Y_\nu=0$. If no
  accretion occurs during the cooling process the transition between the
  different stages  occurs at constant baryonic mass, i.e. along vertical
  lines \cite{prak97}. We identify with a full black dot and a red
  asterisk, respectively, the minimal configuration with a kaon condensate in the
  centar and with a fraction  of at least 10$^-8$ of kaons.

Some  configurations obtained for $S=2$ and
  neutrino-free matter cannot be populated since their baryonic mass is
  larger than the maximum baryonic mass obtained with $S=1$ and trapped
  neutrinos. In particular, all stars with $M>2.05\, M_\odot$ for $S=2$ and
  $L=93$ MeV belong to this set of stars. It is also seen that within the
  QMC model (but not GM1) it may occur that a star has a kaon condensate in its center
  during the neutrino trapped phase. After the neutrinos diffuse out the condensate
  melts and finally it is once more formed  after cooling. These
  transformations should result in neutrino signals after the supernova
  explosion and before the cooling of the star.

 Some conclusions are in order: a) from both Fig. \ref{fig7} and
  Table \ref{table1} we conclude that for $L=93$ MeV no blackhole will be
  formed during the cooling process since the maximum baryonic mass at $S=1$
  and $Y_l=0.4$ is not larger than the  maximum baryonic mass at $S=0$. This
  is not the case for GM1. This model predicts the formation of low mass
  backhole during cooling; b) decreasing the symmetry energy slope $L$
 may  modify some of the above conclusions. In particular, for $L=70$ 
(but not anymore for 59 MeV)
  within QMC there is a small range of star configurations ($\Delta M\sim 0.07
  \ M_\odot$) that will decay to a blackhole during cooling. Within GM1 the
  number of star configurations that decay into blackholes increases when $L$
  goes from 93 to 70 MeV, but decreases if $L$ is further reduced to 59
  MeV. This is due the smaller kaon content in these last stars 
together with a central density
  that is not much larger; c) the cooling of the stars that contain a kaon 
condensate, in
  same cases, involves the melting of the condensate at an intermediate stage
  ($S=2$) and a second formation of the condensate at $T=0$. 

We should point out that we are studying the evolution of the stars without
considering finite size effects as in \cite{maruyama06}. In the present
calculation the kaon potential is not strong enough to give rise to a first
order phase transtion, and corresponds to a second order phase transition.

\section{Conclusions}

In the present work we have revisited the QMC model at finite temperature to 
investigate the thermal kaon effects on stellar properties. The 
$\omega-\rho$ interaction was included because it softens the very hard QMC
symmetry energy at high densities and can be used to tune the values of the 
slope of the symmetry energy. As had already been seen in 
\cite{panda2012,rafael}, lower values of the slope yield  smaller amounts of 
strangeness if hyperons are considered for a given density. 
The same conclusion is here obtained if 
kaons are the carriers of strangeness instead of the hyperons.

As compared with the results obtained with the GM1 parametrization of the NLWM,
the QMC EoS is generally softer, the only exception being the $S=2$ case for
$L=93.5$ MeV (see table \ref{table1}), which is due to the fact that no kaon
condensate is formed because the central  temperatures of the star lie above the
melting temperature of the condensate. A softer EOS at intermediate densities
implies a smaller amount of kaons and as a consequence within QMC no
black-hole formation is expected during the cooling of the a protoneutron
star with a kaon condensate in the core, if $L$ is large enough. For smaller
values of $L$, but not too small,  the set of stars that could cool to a 
black-hole is very reduced
and certainly much smaller than what is expected with GM1. 

It is interesting to
identify the role of the density dependence of the symmetry energy on the
possible evolution of a compact star with a kaon condensate. Within QMC no
black-hole is formed either if $L$ is large or $L$ small.  
This is due to the balance
between the softening of the EOS when $L$ is smaller, together with a less
pronounced softening of the EOS because less kaons are formed. Within GM1 
there are always a quite large set of
stars that cool to a black-hole, although this set is larger for intermediate
values of $L$. 

 We have also shown that the complex evolution of the star may
  include the melting and
formation of a  new  kaon condensate.  The first transformation is driven  by the
neutrino diffusion and  the second is due to  cooling. These processes
could be responsible for a neutrino signal followed by a gamma ray burst after the
supernova explosion. 

Finally, we point out that both models can describe very massive stars, namely stars as massive as the
pulsars PSR J1614–2230 \cite{demorest} and  PSR J0348+0432 \cite{antoniadis}.

\section*{ACKNOWLEDGMENTS}
This work was partially supported by
the initiative 
QREN financed by the UE/FEDER through the Programme COMPETE under 
the project PTDC/FIS/113292/2009,
 by CNPq (Brazil) and FAPESC (Brazil) under 
project 2716/2012,TR 2012000344, and  by NEW COMPSTAR, a COST initiative.
P.K.P. acknowledges the warm hospitality
at both the University of Coimbra and Universidade Federal de Santa Catarina
and D.P.M. at the Universidad de Alicante, where parts of this work were 
carried out.

\end{document}